\documentclass[12pt]{article}
\pdfoutput=1
\usepackage{authblk}
\usepackage{bm}
\usepackage{mathrsfs}
\usepackage{amsbsy}
\usepackage{graphicx}
\usepackage{subfigure}
\usepackage{amsmath}
\usepackage{amssymb}
\usepackage{braket}
\usepackage[numbers,sort&compress]{natbib}
\usepackage{lipsum}
\usepackage[]{hyperref}
\newcommand\blfootnote[1]{
\begingroup
\renewcommand\thefootnote{}\footnote{#1}
\addtocounter{footnote}{-1}
\endgroup
}

\newcommand{\mpref}[1]{Figure.\ref{#1}}
\numberwithin{equation}{section}

\begin{document}
	
	\begin{center}
		{\bf Page Curves for Accelerating Black Holes}\\

		\vspace{1.6cm}
		{\textbf{Ming-Hui Yu}$^{1}$, ~\textbf{Xian-Hui Ge}$^{1}$}\blfootnote{* Corresponding author. gexh@shu.edu.cn}, \textbf{Cheng-Yuan Lu}$^{1}$
		\vspace{0.8cm}
		
		$^1${\it Department of Physics, Shanghai University, Shanghai 200444, China} \\

		\vspace{1.6cm}

		\begin{abstract}
The island paradigm for the fine-grained entropy of Hawking radiation is applied to eternal charged accelerating black holes. In the absence of the island, the entanglement entropy grows linearly and divergent at late times, while once the island outside the event horizon is taken into account, the unitary Page curve is reproduced naturally. The impact of the charge and the acceleration is investigated at late times. For the Page time and the scrambling time, they both increase as the acceleration increases, while decreasing as the charge increases. In particular, neutral black holes have the largest Page time and scrambling time. It is worth noting that the Page time and the scrambling time are divergent in the extremal limit, which implies that islands may be related to the causal structure of spacetime.
		\end{abstract}
	\end{center}
\newpage
\tableofcontents
\newpage

\section{Introduction} \label{introduction}
\qquad Quantum gravity is one of the most active fields in modern physics. The black hole information paradox, which lasted about half a century, is one of the key signposts to the theory of quantum gravity \cite{paradox}. Hawking calculated in 1974 that black holes have a temperature and can emit Hawking radiation, which resembles black-body radiation \cite{HR}. As the black hole evaporates, it takes the information that falls into the black hole with it. Thus, Hawking predicted in 1976 that the evaporation of a black hole would lead to information loss, known as the black hole information paradox \cite{paradox}. This paradox leads to persistent debate and studies in the field of theoretical physics. On the one hand, through quantum mechanics, the evolution of physical systems must follow the principle of the unitary. On the other hand, Hawking radiation is produced using quantum fluctuations near the event horizon of a black hole, so the information loss undermines the unitary. More specifically, a black hole formed from a pure state will evolve into a mixed state during evaporation.
\par The issue has taken a turn for the better until 1993 --- The Page curve \cite{PC1,PC2}, which describes the entanglement entropy of Hawking radiation is depicted in detail by the Page theorem \cite{Page theorem} under the unitary hypothesis. The black hole information paradox can be solved if the Page curve can be calculated directly from semi-classical gravity without this hypothesis. Although still difficult, the development of the holographic principle provides confidence in this issue \cite{adscft}.
The AdS/CFT duality has been an important and fascinating topic, not only providing a key link between the theory of gravity in AdS  (anti-de Sitter) spacetime and the CFT (conformal field theory) that lives at its boundary, but also opening a window into quantum gravity. Many recent advances in holographic descriptions of the evaporation of black holes further support the view of information conservation. The Ryu-Takayanagi (RT) formula for calculating the holographic entanglement entropy is undoubtedly the best of these works \cite{RT,QRT}.  When one considers higher-order quantum corrections, the RT formula is generalized to the quantum extremal surface (QES) formula of the entropy functional of the sum of the contributions of the bulk field and the area term \cite{QES}. Once we use the QES formula to calculate the entropy of Hawking radiation based on the semi-classical gravity method, the Page curve is naturally reproduced \cite{bulk entropy}. In particular, the most critical point is that the bulk field contribution contains a disconnected region located in the interior of the black hole, called the ``\emph{island}'' \cite{island rule}. The idea of reconstructing the Page curve by correctly tracking the radiation outside black holes and the interior of black holes is summarized as the ``\emph{entanglement wedge reconstruction}'' \cite{entanglement wedge}. Finally, we can express the formula for calculating the fine-grained entropy of Hawking radiation as the so-called ``\emph{island formula}'' \cite{review}:
\begin{equation}
S_{\text{Rad}}=\text{Min}\big[\text{Ext}(S_{\text{gen}})\big],  \label{island foumula1}
\end{equation}
where $S_{\text{gen}}$ represents the generalized entropy of radiation, which can be written as
\begin{equation}
S_{\text{gen}}=\frac{\text{Area}(\partial I)}{4G_N} +S_{\text{bulk}} (I \cup U), \label{generalized entropy}
\end{equation}
where the first term is the area term of islands, in which $I$ refer to the island, $\partial I$ is denoted the boundary of the island. The second term is the entropy that is contributed by matter fields around the black hole. One should note that besides the radiation region $R$, the region $I$ inside the black hole is also considered. It also provides a glimmer of insight into the interior of black holes. In brief, the island formula \eqref{island foumula1} gives us an instruction: in the explicit calculation, one first extremize the generalized entropy \eqref{generalized entropy}, these saddle points corresponding to positions of QES, namely the island. Then, the minimal value of the entropy is our candidate. For the whole process of an evaporating black hole, at early times, the QES is just a trivial (vanishing) surface, so there is no island at early times, which leads to a Hawking curve. However, the minimal condition requests that the QES is a non-trivial (non-vanishing) surface at late times, which provides a decreasing curve for the entropy. Therefore, combining two curves, we finally obtain the Page curve. The island formula \eqref{island foumula1} then can be rewritten as follows:
\begin{equation}
S_{\text{Rad}}=\text{Min} \big[S_{\text{gen}} (\text{without island}),S_{\text{gen}} (\text{with island})\big].
\end{equation}
\par The first fruits of victory were harvested in the evaporating Jackiw--Teitelboim (JT) black hole in two-dimensional (2D) gravity, where the auxiliary thermal bath is coupled to the asymptotically AdS black hole \cite{bulk entropy,eternal bh}. However, the range of adaptation of the island paradigm extends far beyond the framework of AdS black holes, one can refer to \cite{jt1,jt2,2d1,2d2,2d3,2d4,2d5,2d6,2d7,2d8,2d9,high1,high2,high3,high4,high5,high6,high7,high8,high9,high10,high11,high12,high13,high14,high15,high16,high17,high18,high19,high20,high21,high22,high23,high24,high25,high26,high27,high28,high29,high30,high31,high32,high33,high34,high35,high36,high37,high38,high39,high40,high41,hyper,inf1,inf2,high42,high43,high44,high45} for an inexhaustive list of recent studies.
\par There is no doubt that 2D gravity, despite the simplicity of calculations it provides, is still only a toy model. The more representative and acceptable Einstein gravity, in which the Kerr-Newmann family is a typical four-dimensional (4D) black hole prototype. They are described by the mass, charge, and angular momentum parameters. However, there is a little-known exact solution for a 4D black hole: the ``$C$-metric'', which describes an accelerating black hole \cite{cmetric1,cmetric2,cmetric3}. For this kind of interesting black hole, there is a conical defect angle along the polar axis of the black hole. It is because of the defect angle that it provides the driving force for the black hole to accelerate. The exact solution described by the ``$C$-metric'' is ideal. Nevertheless, its applications are widespread beyond the theory of classical general relativity. It can also describe the generation of black hole pairs in electromagnetic fields \cite{pair}, and the splitting of cosmic strings \cite{cosmic}, especially recent work on its thermodynamics \cite{thermodynamics}. More work in this field can be referred to \cite{accelerating btz,pv1,pv2}
\par Based on these remarkable special properties of the $C$-metric, in this paper we investigate the information issue in the charged version of the $C$-metric. In order to solve the information issue, we calculate the entanglement entropy of the charged accelerating black hole explicitly using the island formula \eqref{island foumula1} and reconstruct the corresponding Page curve. We also discuss the effects of the charge and the acceleration, in particular, for the neutral black hole, which has the largest Page time. In addition, it is noted that in the extremal case, due to the vanishing of the temperature, some physical quantities become divergent, which strongly implies that quantum information is closely related to the causal structure of spacetime.
\par The structure of this paper is as follows. In Sec.\ref{Review for Accelerating Black Holes}, we briefly review some properties of the charged $C$-metric. In Sec.\ref{Island Paradigm for Black Holes}, the entanglement entropy of Hawking radiation in the with and without island configuration is calculated by the island paradigm. In Sec.\ref{Page Curve and Scrambling Time}, we obtain the Page curve satisfying the unitary and the scrambling time, then discuss the impact of the charge and the acceleration parameters. The summary is in Sec.\ref{Summary}. In what follows, we will use the nature units where $c=\hbar=k_B=1$.

\section{Review for Accelerating Black Holes} \label{Review for Accelerating Black Holes}
\qquad In this section, we give a brief review of accelerating black holes.  Physically, the accelerating black holes can be produced by a cosmic string collision with a black hole, so that a gravitational interaction could result in an acceleration of the black holes. The accelerating black hole can be described by the $C$-metric, which represents a pair of black holes that accelerating away from each other in an opposite direction. For the charged version of the $C$-metric, it can be regarded simply as a generalization of Reissner-Nordstr$\ddot{\text{o}}$m (RN) black holes. Besides the parameters of mass $M$ and electric charge $Q$, it also includes an extra parameter which associated with the acceleration $\alpha$. This charged $C$-metric can be written in a spherical coordinate as follows \cite{cmetric}
\begin{equation}
ds^2=\frac{1}{\Omega^2} \bigg[-f(r)dt^2+\frac{dr^2}{f(r)}+r^2\bigg(\frac{d\theta^2}{P(\theta)} + P(\theta) \sin^2 \theta d\phi^2 \bigg) \bigg],  \label{metric1}
\end{equation}
where
\begin{subequations}
\begin{align}
f(r)     &=\big(1-\alpha^2r^2\big)\bigg(1-\frac{2M}{r}+\frac{Q^2}{r^2} \bigg),  \label{metric function} \\
P(\theta)&=1+2\alpha M \cos \theta +\alpha^2 Q^2 \cos^2 \theta,   \label{angular function} \\
\Omega   &=1+\alpha r \cos \theta.  \label{conformal factor}
\end{align}
\end{subequations}
The gauge potential is given by
\begin{equation}
F=dA, \qquad A=-\frac{Q}{r}dt.  \label{gauge}
\end{equation}
One can easily finds that this metric is asymptotically flat. In the limit $\alpha \to 0$, the charged $C$-metric approaches the RN black hole. As for the main features of accelerating black holes, one of the key features is that they would be moving through space at a constant velocity rather than remaining stationary. Another important property of black holes is that these accelerating black holes would emit gravitational waves. Further, in the vanishing charge case, one can obtain Schwarzschild black holes. There is an intrinsic singularity at $r=0$, but there exist three coordinate singularities at the root of the metric function $f(r)$, then we define the following constraints:
\begin{subequations}
\begin{align}
r_{\alpha}&=\frac{1}{\alpha}, \label{accelerating horizon} \\
r_+       &=M+\sqrt{M^2-Q^2}, \label{event horzion} \\
r_-       &=M-\sqrt{M^2-Q^2}, \label{cauchy horzion}
\end{align}
\end{subequations}
which corresponds to the acceleration horizon, the event horizon, and the Cauchy horizon, respectively. In order to avoid the naked singularity, one must insist on the following relation:
\begin{equation}
r_- \le r_+ \le r_{\alpha}. \label{nake}
\end{equation}
Thus, for non-extremal case, namely $r_+<r_-<r_{\alpha}$, the metric function $f(r)$ is always positive, which also imply that the angular function $P(\theta)$ is also positive for the range $[0,\pi]$. Different from RN black holes, in addition to become extremal when $M=Q$, i.e. the event and Cauchy horizons are coincides. The accelerating black hole can also becomes extremal for $\alpha=\frac{1}{r_+}$, which is called the Nariai limit. In this paper, we mainly consider the non-extremal case for simplicity.
\par The other interesting property of accelerating black holes is that there are two conical singularities at the axis at $\theta=0$ and $\theta=\frac{\pi}{2}$. Therefore, the ratio of the circumference to the radius there is not to equal to $2\pi$. Nevertheless, one can still remove one of two conical singularities by resetting the coordinate $\phi$. More specifically, we first consider a range for $\phi \in [0, 2\pi a)$, where $a$ is a constant. Then, if there is regular at $\theta=0$ with a constant $t$ and $r$, we obtain
\begin{equation}
\frac{C_m}{R}=\lim_{\theta \to 0} \frac{2\pi a P(\theta) \sin \theta}{\theta}=2\pi a P(0), \label{circumference1}
\end{equation}
where $C_m$ and $R$ stand for ``circumference'' and ``radius'', respectively. Similarity, if the regularity is insisted at $\theta=\pi$, then
\begin{equation}
\frac{C_m}{R}=\lim_{\theta \to \pi} \frac{2\pi a P(\theta) \sin \theta}{\pi - \theta}=2\pi a P(\pi), \label{circumference2}
\end{equation}
Thus, the equations \eqref{circumference1} and \eqref{circumference2} indicate that there are different conical singularities for different conicities. One of the two conical singularities can be eliminated, but not both when we choose a reasonable constant $a$. For example, if one selects that $a=P^{-1}(0)$, then the deficit angles at the poles $\theta=0$ and $\theta=\pi$ are \cite{cmetric}
\begin{equation}
\delta(0)=0, \qquad \delta(\pi)=2\pi \bigg[1-\frac{P(\pi)}{P(0)} \bigg]. \label{deficit angle1}
\end{equation}
In this way, we remove the conical singularity at $\theta=0$ and left a deficit angle at $\theta=\pi$. Likewise, if we choose
$a=P^{-1}(\pi)$, which leads to a deficit angle at $\theta=0$ and the regularity at $\theta=\pi$ is maintained, i.e.
\begin{equation}
\delta(\pi)=0, \qquad \delta(0)=2\pi \bigg[1-\frac{P(0)}{P(\pi)} \bigg]. \label{deficit angle2}
\end{equation}
Henceforth, we choose $a=P^{-1}(0)$ and preserve the equation \eqref{deficit angle1} for simplicity.
\par Now, we introduce some thermodynamic qualities. For the Hawking temperature, we can use the standard Euclidean method: we analytically continue to the Euclidean signature, namely
\begin{equation}
t \to -i \tau, \qquad \tau \sim \tau+\beta, \label{imaginary time}
\end{equation}
with $\beta$ is the inverse temperature. Then the temperature of black holes is determined by
\begin{equation}
\begin{split}
T_H=T \big |_{r=r_+}&=\frac{(r_+-r_-)(1-\alpha^2 r_+^2)}{4 \pi r_+^2}  \\
   &=\frac{\sqrt{M^2-Q^2} \bigg(1- \alpha^2 \big( M+ \sqrt{M^2-Q^2} \big)^2 \bigg) }{2\pi \big(M+ \sqrt{M^2 -Q^2} \big)^2}. \label{temperature}
\end{split}
\end{equation}
Meanwhile, one can determine the Bekenstein-Hawking entropy, which is a quarter of the horizon area $\cal{A}$ \cite{thermodynamics}
\begin{equation}
\begin{split}
S_{BH}&=\frac{1}{4G_N}\int_{-\pi a}^{\pi a} \int_0^{\pi} \sqrt{g_{\theta \theta}  g_{\phi \phi}} \Bigg |_{r=r_+} d\theta d\phi =\frac{\pi r_+^2}{G_N} \cdot \frac{1}{(1-\alpha^2 r_+^2) P(0)} \\
&=\frac{\pi \big[M+\sqrt{M^2-Q^2} \Big]^2}{G_N \Big( 1+2M\alpha+\alpha^2 Q^2 \Big)\Big[1- \Big(M+\sqrt{M^2-Q^2} \Big)^2 \alpha^2 \Big]}. \label{bh entropy}
\end{split}
\end{equation}
This clearly differs from the RN black hole in which the extremal case corresponds to the finite entropy. This means that entropy is not well-defined in the Nariai limit. However, the entropy is finite in the charged limit. Finally, we perform the Kruskal transformation to obtain the maximum extended spacetime. For the charged $C$-metric \eqref{metric1}, we define the tortoise coordinate as
\begin{equation}
\begin{split}
r_{\star}&=\int \frac{1}{f(r)}dr=\frac{(r_+-r_- + \alpha^2 r_+^2 r_- -\alpha^2 r_+ r_-^2) \log \big | \frac{1+\alpha r}{1- \alpha r} \big |}{2\alpha (r_+-r_-)(\alpha^2 r_+^2 -1)(\alpha^2 r_-^2 -1)}\\
&+\frac{2r_-^2 (\alpha^2 r_+^2-1) \log |r-r_-| - 2r_+^2 (\alpha^2 r_-^2 -1) \log |r-r_+| - (r_+^2 -r_-^2) \log |1- \alpha^2 r^2|}{2(r_+-r_-)(\alpha^2 r_+^2 -1)(\alpha^2 r_-^2 -1)}. \label{tortoise}
\end{split}
\end{equation}
So the null Kruskal coordinates are\footnote{This expression is only valid for the right wedge of the spacetime. We should swap the signs of $U$ and $V$ when we calculate for the left wedge.}
\begin{equation}
\begin{split}
U(t,r_{\star}) &= - e^{-\kappa_+ u}=- e^{-\kappa_+(t-r_{\star})} ,\\
V(t,r_{\star}) &= + e^{+\kappa_+ v}= + e^{+\kappa_+(t+r_{\star})}, \label{kruskal}
\end{split}
\end{equation}
where $\kappa_+ =2 \pi T_H$ is the surface gravity on the event horizon. Thus, the metric \eqref{metric1} can be recast as
\begin{equation}
d\tilde{s}^2=-\frac{dU dV}{g^2(r)}+\frac{r^2}{\Omega^2}  \bigg( \frac{d \theta^2}{P(\theta)} +P(\theta) \sin^2 \theta d\phi^2 \bigg), \label{metric2}
\end{equation}
with the conformal factor
\begin{equation}
\begin{split}
g^2(r)&=\frac{k_+^2 e^{2 \kappa_+ r_{\star}(r)}}{f(r)}\\
      &=\frac{(r_+-r_-)^2}{4r_+^4} e^{- \frac{\frac{1}{2} (r_+ -r_-) (1+\alpha^2 r_+ r_-) \log \big| \frac{1+ \alpha r}{1-\alpha r} \big| }{\alpha r_+^2 (\alpha^2 r_-^2 -1)}} r^2 (r-r_-)^{\frac{r_+^2 +r_-^2 - 2\alpha^2 r_+^2 r_-^2 }{r_+^2 (\alpha^2 r_-^2 -1)}} \\
      & \times (\alpha^2 r_+^2 -1)^2 (1- \alpha^2 r^2)^{\frac{3r_+^2-r_-^2-2\alpha^2 r_+^2 r_-^2}{2r_+^2 (\alpha^2 r_-^2 -1)}}. \label{conformal factor2}
\end{split}
\end{equation}
Here, we have rescaled the metric \eqref{metric1} conformally by $\Omega$ \eqref{conformal factor}: $d\tilde{s}^2=\Omega^2 ds^2$. Such rescaling would provide a simpler form but not affect the physical results in the following section.

\section{Island Paradigm for Black Holes} \label{Island Paradigm for Black Holes}
\par In this section, we investigate the entanglement entropy between the radiation and black holes, together with the inclusion of the island. We start with the no-island case, then consider the effect of islands.
\subsection{Early times: Without Island}
\par
\begin{figure}[htb]
\centering
\includegraphics[scale=0.20]{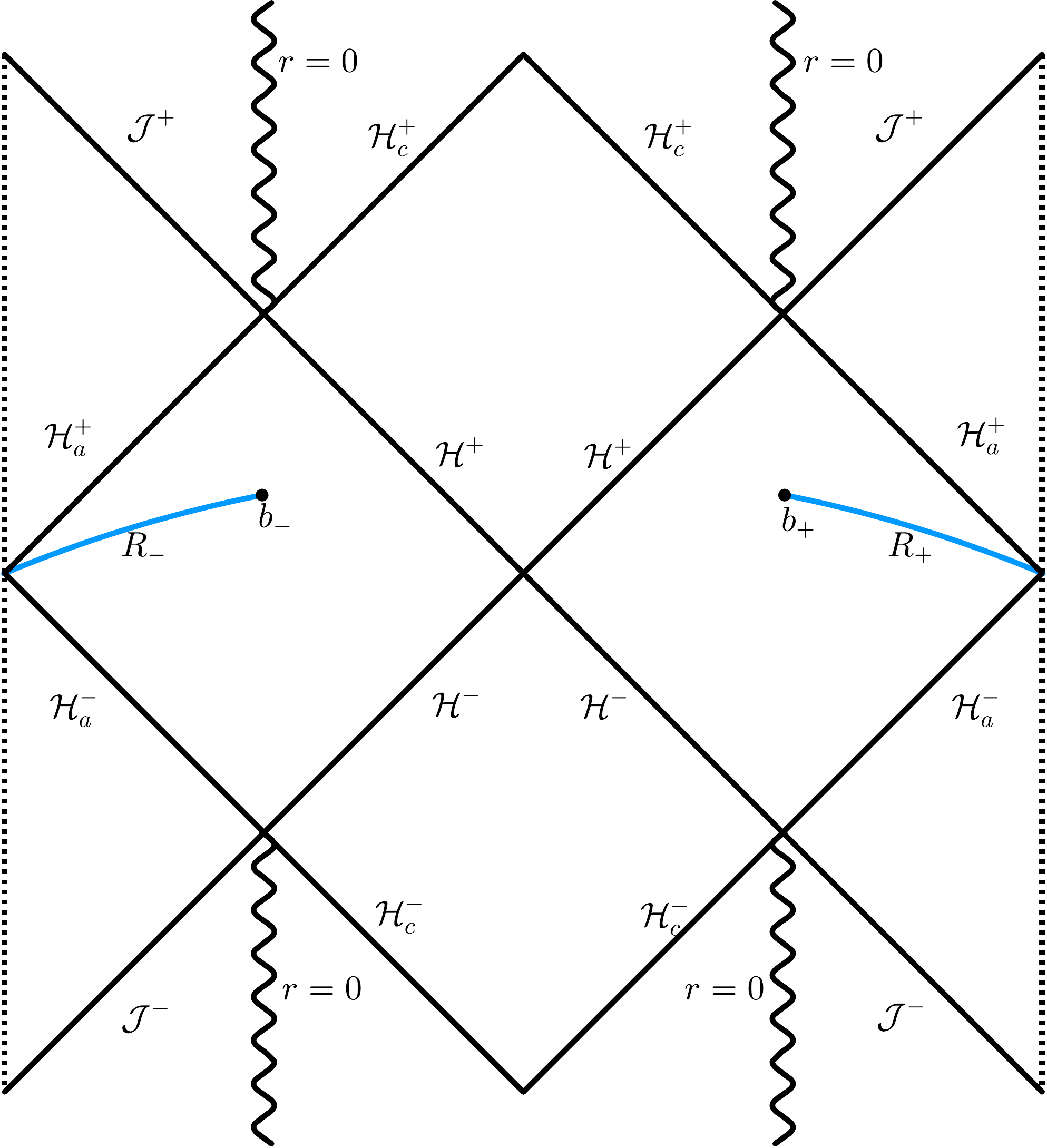}
\caption{The Penrose diagram for the charged $C$-metric. The solid lines represents null hypersurfaces. The dashed lines are identity. $\cal{H}^{\pm}$, ${\cal{H}}_c^{\pm}$, and ${\cal{H}}_{a}^{\pm}$ represents the future/past event horizons, acceleration horizons, and Cauchy horizons, respectively. Blue lines are labeled as radiation regions $R_{\pm}$, where the boundaries are noted by $b_{\pm}=(\pm t_b,b)$ for the left and right wedge. The singularity is located at $r=0$.}
\label{penrose1}
\end{figure}

 The total system is a thermofield double state, or equivalently, the Hartle-Hawking state, since we consider an eternal black hole. We first assume that the black hole is formed by a pure state and then make an $S$-wave approximation. Under this assumption, we can ignore the angular directions $\theta$ and $\phi$, and only focus on the contribution of the radial direction. Thus, the metric \eqref{metric2} is reduced to a 2D form:
\begin{equation}
ds^2=-\frac{1}{g^2(r)} dU dV.  \label{metric3}
\end{equation}
The corresponding Penrose diagram is plotted in \mpref{penrose1}. In the absence of island, there is only the radiation part left in the island formula \eqref{island foumula1}. For a thermofield double state
\begin{equation}
\ket{\text{TFD}}=\frac{1}{\sqrt{Z}} \sum_n e^{\frac{\beta E_n}{2}} \ket{n}_L \otimes \ket{n}_R. \label{tfd}
\end{equation}
Thus the relation $\big(H_L -H_R \big) \ket{\text{TFD}}=0$ is maintained for such a state and these two subsystems are maximally entangled. To sum up, the mutual information between the radiation at the left and the right wedge can be calculated by 2D CFT \cite{EE formula}
\begin{equation}
I(R_-,R_+)=S_{\text{bulk}}(R)=\frac{c}{3} \log d(b_+,b_-),  \label{mutual information}
\end{equation}
where $c$ is the central charge, $d(b_+,b_-)$ is the geodesic distance between the points $b_+$ and $b_-$. For the metric \eqref{metric3}, the geodesic distance $d(b_+,b_-)$ is represented by
\begin{equation}
\begin{split}
d(b_+,b_-)&=\sqrt{-ds^2}\\
          &=\sqrt{-\frac{[U(b_+)-U(b_-)][V(b_+)-V(b_-)]}{g(b_+)g(b_-)}}. \label{geodesic}
\end{split}
\end{equation}
We substitute the coordinate of $b_{\pm}$, then obtain the result as
\begin{equation}
\begin{split}
S_{\text{bulk}}&=\frac{c}{6} \log \bigg[ \frac{1}{g^2(b)} 2e^{2 \kappa_+ r_{\star}(b)} [2\cosh(2 \kappa_+ t_b)+1] \bigg]
                =\frac{c}{6} \log \bigg[ \frac{4f(b)}{\kappa_+^2} \cosh^2 (\kappa_+ t_b) \bigg] \\
               &=\frac{c}{6} \log \Bigg[ \frac{16r_+^4 (r_+-b)(b-r_-)(\alpha^2b^2-1)}{(r_+-r_-)^2 b^2 (\alpha^2 r_+^2-1)^2  } \cosh^2 (\kappa_+ t_b) \Bigg].
\end{split}
\end{equation}
At late times limit, $t \gg \kappa_+$, we can use the approximation: $\cosh(\kappa_+ t_b) \simeq \frac{1}{2} e^{\kappa_+ t_b}$, then
\begin{equation}
\begin{split}
S_{\text{bulk}} (R) &\simeq \frac{c}{3} \log [\cosh (\kappa_+ t_b)] \\
                    &\simeq \frac{c}{3} \kappa_+ t_b = \frac{2c}{3} \pi T_H t_b \\
                    & =\frac{2c \pi t_b}{3}  \frac{\sqrt{M^2-Q^2} \big[1- \alpha^2 \big( M+ \sqrt{M^2-Q^2} \big)^2 \big] }{2\pi \big(M+ \sqrt{M^2 -Q^2} \big)^2}.  \label{result1}
\end{split}
\end{equation}
\par One can obviously see from this result that the entanglement entropy of radiation is proportional to the time $t_b$ for the observer at infinity and the temperature of black holes. Besides, we also find that increasing acceleration will decrease the rate of growth. Finally, the entanglement entropy behaves divergent at late times. In particular, for an eternal black hole, the amount of radiation is infinite at late times, which leads to the entanglement entropy far exceeding the Bekenstein-Hawking entropy bound \cite{entropy bound}. Therefore, there exists an intuitive paradox here: on the one hand, our result provides a linear increase in the entanglement entropy; On the other hand, the fine-grained (entanglement) entropy must be small than the coarse-grained (Bekenstein-Hawking) entropy. We expect to solve this thorny issue through the island paradigm.

\subsection{Late times: With Island}
\par
\begin{figure}[htb]
\centering
\includegraphics[scale=0.20]{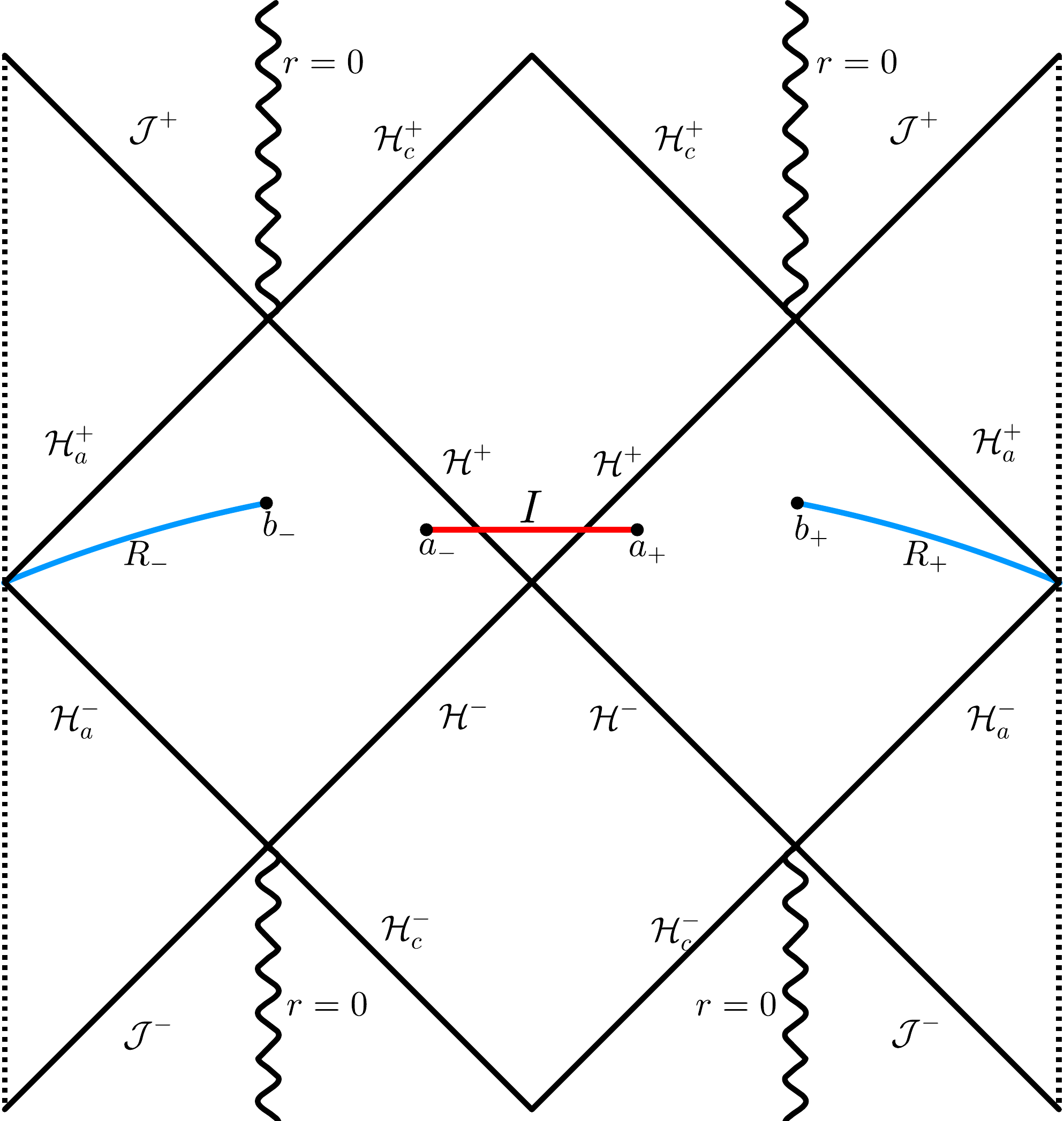}
\caption{The Penrose diagram for charged accelerating black holes with an island. The disconnected island region is labeled as $I$, its boundary is $a_{\pm}=(\pm t_a,a)$.}
\label{penrose2}
\end{figure}

 Now we introduce the disconnected island region in our system by following the island paradigm. The Penrose diagram is shown in \mpref{penrose2}.
At this time, the von Neuman entropy of the matter part is given by \cite{EE formula}
\begin{equation}
S_{\text{bulk}}(I \cup R) =\frac{c}{6} \log  \Bigg[ \frac{d(a_+,a_-)d(b_+,b_-)d(a_+,b_+)d(a_-,b_-)}{d(a_+,b_-)d(a_-,b_+)}  \Bigg].
\end{equation}
However, we still assume that the late times and the large distance are allowed. This approximation allows us to rewrite the above equation into a simpler form
\begin{equation}
d(a_+,a_-) \simeq d(b_+,b_-) \simeq d(a_+,b_+)\simeq d(a_-,b_+) \gg d(a_+,b_+) \simeq d(a_-,b_-),
\end{equation}
and
\begin{equation}
\begin{split}
S_{\text{bulk}} &\simeq \frac{c}{6} \log [d(a_-,b_-) d(a_+,b_+)] \\
                &=\frac{c}{3} \log \Bigg[ \frac{2e^{\kappa_+ (r_{\star}(a)+r_{\star}(b))}}{g(a) g(b)}  \bigg[ \cosh [\kappa_+ (r_{\star}(a)-r_{\star}(b))] -\cosh[\kappa_+ (t_a-t_b)] \bigg] \Bigg].
\end{split}
\end{equation}
Accordingly, the generalized entropy \eqref{generalized entropy} read as
\begin{equation}
\begin{split}
S_{\text{gen}}&=\frac{2\pi a^2}{G_N} \frac{1}{(1-\alpha^2 a^2 ) \big[ 1+\alpha (r_++r_-) +\alpha^2 r_+r_- \big]} \\
              &+\frac{c}{6} \log \Bigg[ \frac{4f(a)f(b)}{\kappa_+^4}  \Big( \cosh[\kappa_+ (r_{\star}(a) - r_{\star}(b) )] -\cosh[\kappa_+ (t_a-t_b)]    \Big)^2 \Bigg]   .  \label{generalized entropy2}
\end{split}
\end{equation}
Extremizing the above expression with respect to $t_a$ firstly
\begin{equation}
\frac{S_{\text{gen}}}{\partial t_a} =\frac{c}{3} \frac{- \kappa_+ \sinh[\kappa_+ (t_a-t_b)]}{\cosh \big[\kappa_+ (r^{\star}(a)-r^{\star}(b) \big]-\cosh[\kappa_+(t_a-t_b)]}=0.   \label{respect time}
\end{equation}
Thus, the only solution for the equation is $t_a=t_b$. Substituting this relation and then extremizing it with respect to $a$:
\begin{equation}
\begin{split}
\frac{S_{\text{gen}}}{\partial a}&=\frac{4 \pi a}{G_N (1+\alpha r_+)(1+\alpha r_-) (\alpha^2 a^2 -1)^2}\\
                                 &+\frac{c \bigg( f^{\prime}(a) -2 \kappa_+ \frac{e^{\kappa_+ x }+1}{e^{\kappa_+ x }-1}   \bigg) }{6f(a)}=0,  \label{respect space1}
\end{split}
\end{equation}
Where $x \equiv r_{\star}(b) -r_{\star}(a)$, and $f^{\prime}(r)$ is the derivative of $f(r)$. Now, we make the near horizon limit, where $a \simeq r_+$. By taking this approximation, the above equation is recast as
\begin{equation}
\begin{split}
\frac{S_{\text{gen}}}{\partial a} \simeq \frac{4 \pi}{G_N \alpha^6 a^3 r_+ r_-} + \frac{c}{6}  \frac{2e^{\kappa_+ (r_{\star}(b) -r_{\star}(a)  )}}{(e^{\kappa_+ (r_{\star}(b) - r_{\star}(a))}-1)(a-r_+)}   =0. \label{respect space2}
\end{split}
\end{equation}
Thus, we finally obtain the location of the quantum extremal island is
\begin{equation}
\begin{split}
a  &\simeq r_+ + \frac{c G_N  \alpha^6 r_+^4 r_- }{12 \pi e^{\kappa_+ r_{\star}(b)} } \\
   &\simeq r_+ + {\cal O} \Big( \frac{cG_N}{r_+} \Big).  \label{location}
 \end{split}
\end{equation}
Therefore, at late times, the entanglement entropy of Hawking radiation with the island is given by
\begin{equation}
\begin{split}
S_{\text{Rad}} &= \frac{2 \pi r_+^2}{G_N}  \frac{1}{(1-\alpha^2 r_+^2) [1+ \alpha (r_++r_-) + \alpha^2 r_+r_-]} \\
               & + \frac{c}{3} \log \Bigg[ \frac{cG_N r_+^2 (b-r_+)(b-r_-) (r_+-r_-)(\alpha^2 b^2 -1)(\alpha^2 r_+^2 -1)}{(r_+-r_-)(\alpha^2 r_+^2 -1)r_+ b^2}  \Bigg] \\
               &\simeq 2S_{BH}.  \label{result2}
\end{split}
\end{equation}
Therefore, we find that once we consider the contribution of an island, the fine-grained entropy of radiation approaches a saturated value eventually. This is a surprising result which is the opposite of Hawking's calculation: at late times, the entanglement entropy of radiation stops increasing and tends to be about twice of the Bekenstein-Hawking entropy. This conclusion also respects the principle of unitary. Now the black hole has enough degree of freedom to entangle with the external radiation, since now the fine-grained entropy of black holes is smaller than the coarse-grained/Bekenstein-Hawking entropy. Therefore, we obtain the expression of the fine-grained entropy of Hawking radiation for the whole process of the evaporation
\begin{equation}
S_{\text{FG}}=\text{Min} \bigg[ \frac{2c \pi}{3} T\cdot t ,\  2S_{\text{BH}} \bigg], \label{final result}
\end{equation}
where the effect of charges and acceleration is absorbed in the temperature $T_H$ and the Bekenstein-Hawking entropy $S_{BH}$, which we will discuss in detail below.

\section{Page Curve and Scrambling Time} \label{Page Curve and Scrambling Time}
\par At present, we have calculated the behavior of the entanglement entropy of radiation from the discussion of applying the island paradigm in the previous section. Therefore, the natural thing to do in this section is to plot the corresponding Page curve and obtain some by-products, such as the scrambling time of black holes.
\par
\begin{figure}[htb]
\centering
\subfigure[\scriptsize{}]{\label{page1}
\includegraphics[scale=0.4]{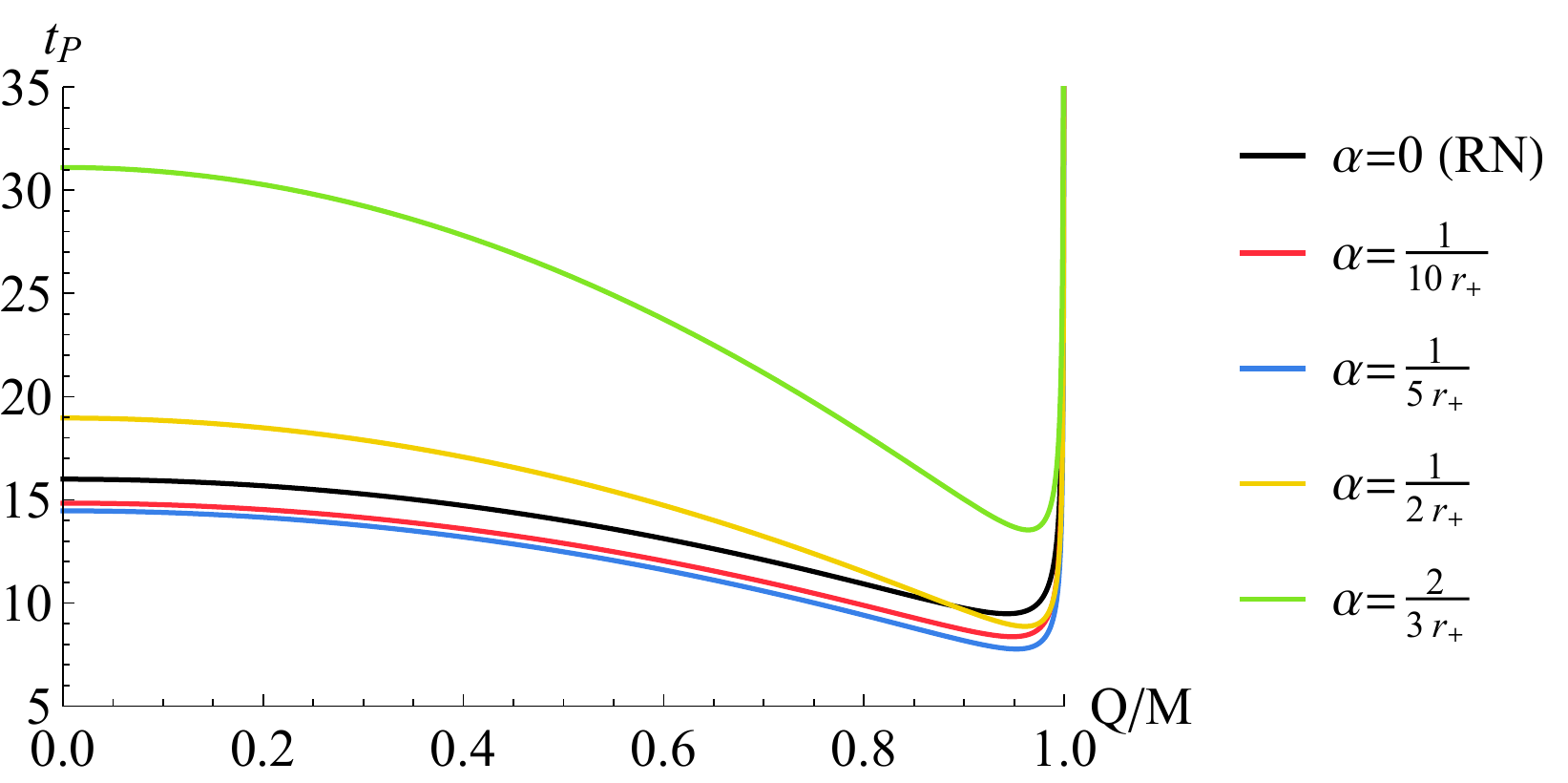}
}
\quad
\subfigure[\scriptsize{}]{\label{page2}
\includegraphics[scale=0.4]{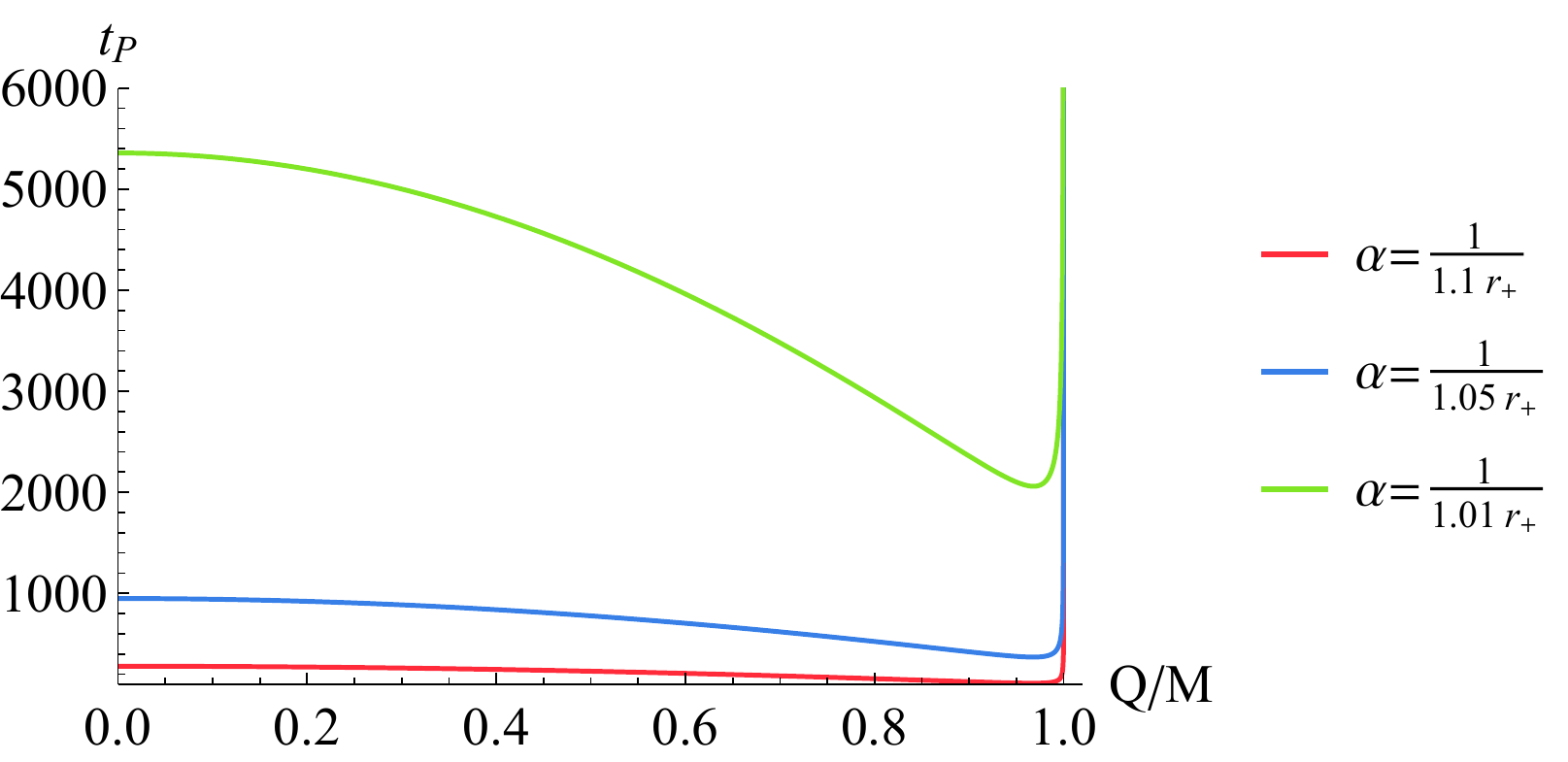}
}
\caption{The Page time by charged accelerating black holes as the function for charges $Q$, in which the acceleration $\alpha$ is fixed (in the unit of $\frac{6 \pi}{c G_N}$ and set $M=1$). On the left, it is the non-extremal case with a small amount of charges, while on the right is the near extremal case.}
\label{pagetime1}
\end{figure}

\begin{figure}[htb]
\centering
\subfigure[\scriptsize{}]{\label{page3}
\includegraphics[scale=0.4]{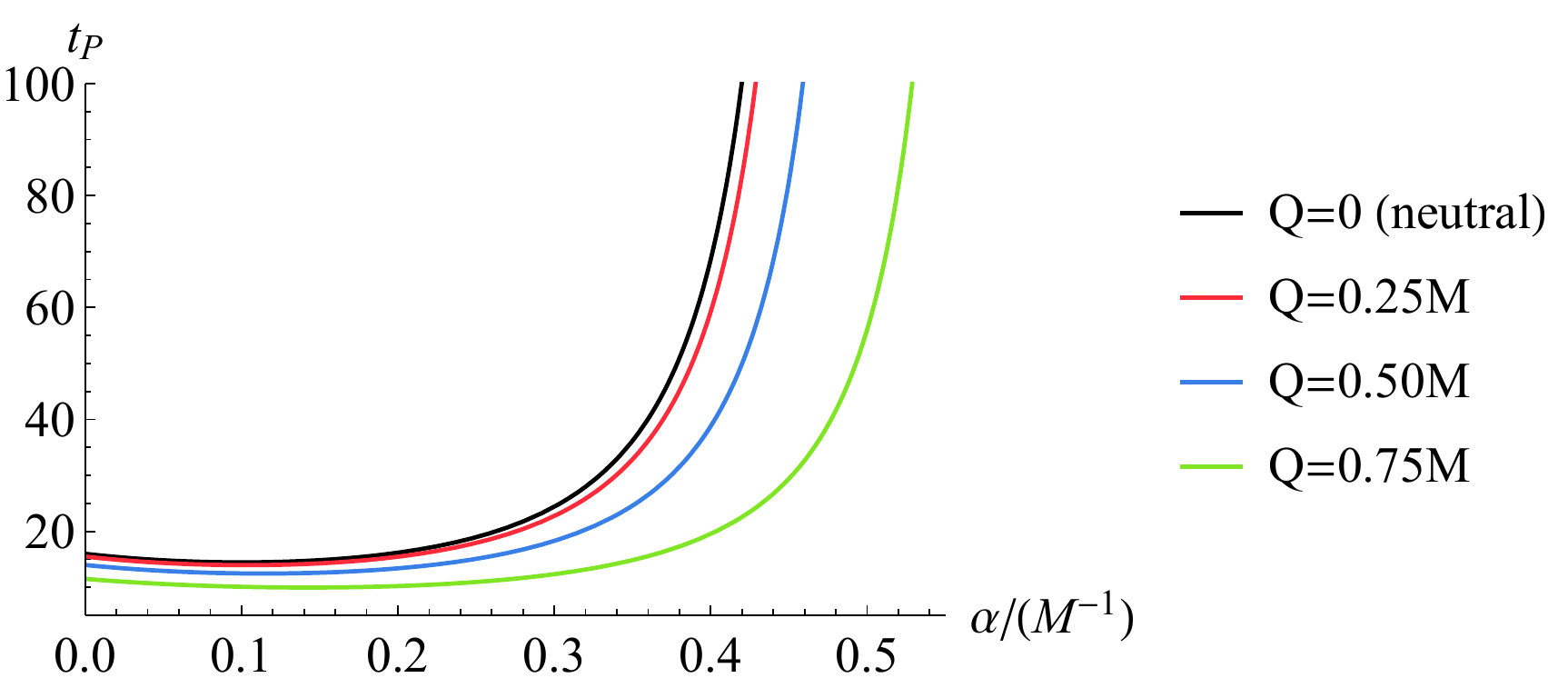}
}
\quad
\subfigure[\scriptsize{}]{\label{page3}
\includegraphics[scale=0.55]{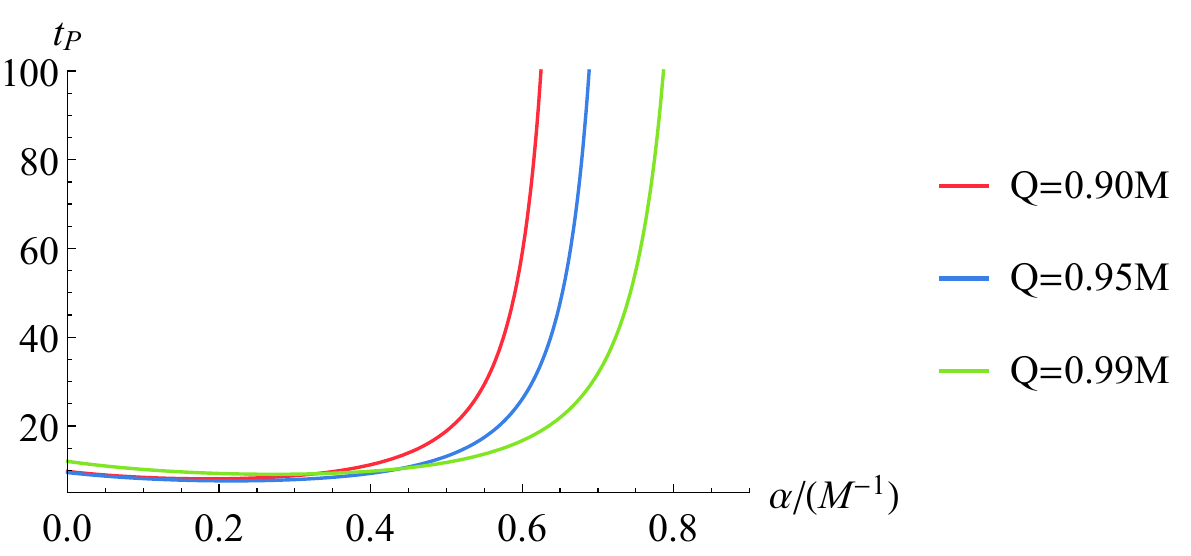}
}
\caption{The Page time by charged accelerating black holes as the function for the acceleration $\alpha$ (in the unit of $\frac{6 \pi}{c G_N}$ and set $M=1$). On the left is the non-extremal case, while on the right is the near extremal case.}
\label{pagetime2}
\end{figure}

 At first, we derive the Page time that defined as the moment when the entanglement entropy approaches the peak. In the language of the entanglement island, the entropy without island \eqref{result1} is roughly equal to the entropy with island \eqref{result2} at the Page time. Accordingly, we obtain the Page time is
\begin{equation}
\begin{split}
t_{\text{Page}} &= \frac{3 S_{BH}}{\pi c T_H} = \frac{12 \pi r_+^4}{ cG_N (r_+-r_-) (\alpha r_+ -1)^2 (\alpha r_+ +1)^3 (\alpha r_- +1) }  \\
&=\frac{6 \pi \big(M+ \sqrt{M^2-Q^2} \big)^4 }{cG_N \sqrt{M^2-Q^2}  \bigg(1+M \alpha - \alpha \sqrt{M^2-Q^2} \bigg) \bigg[ (M+\sqrt{M^2 -Q^2}) \alpha -1 \bigg ]^2  \bigg[(M+\sqrt{M^2 -Q^2}) \alpha+1 \bigg]^3 }.  \label{tp}
\end{split}
\end{equation}
The corresponding graphs as a function for the charge $Q$ and the acceleration $\alpha$ are shown in \mpref{pagetime1}($\alpha$ is fixed) and \mpref{pagetime2}($Q$ is fixed), respectively.
\par
\begin{figure}[htb]
\centering
\subfigure[\scriptsize{}]{\label{pc1}
\includegraphics[scale=0.4]{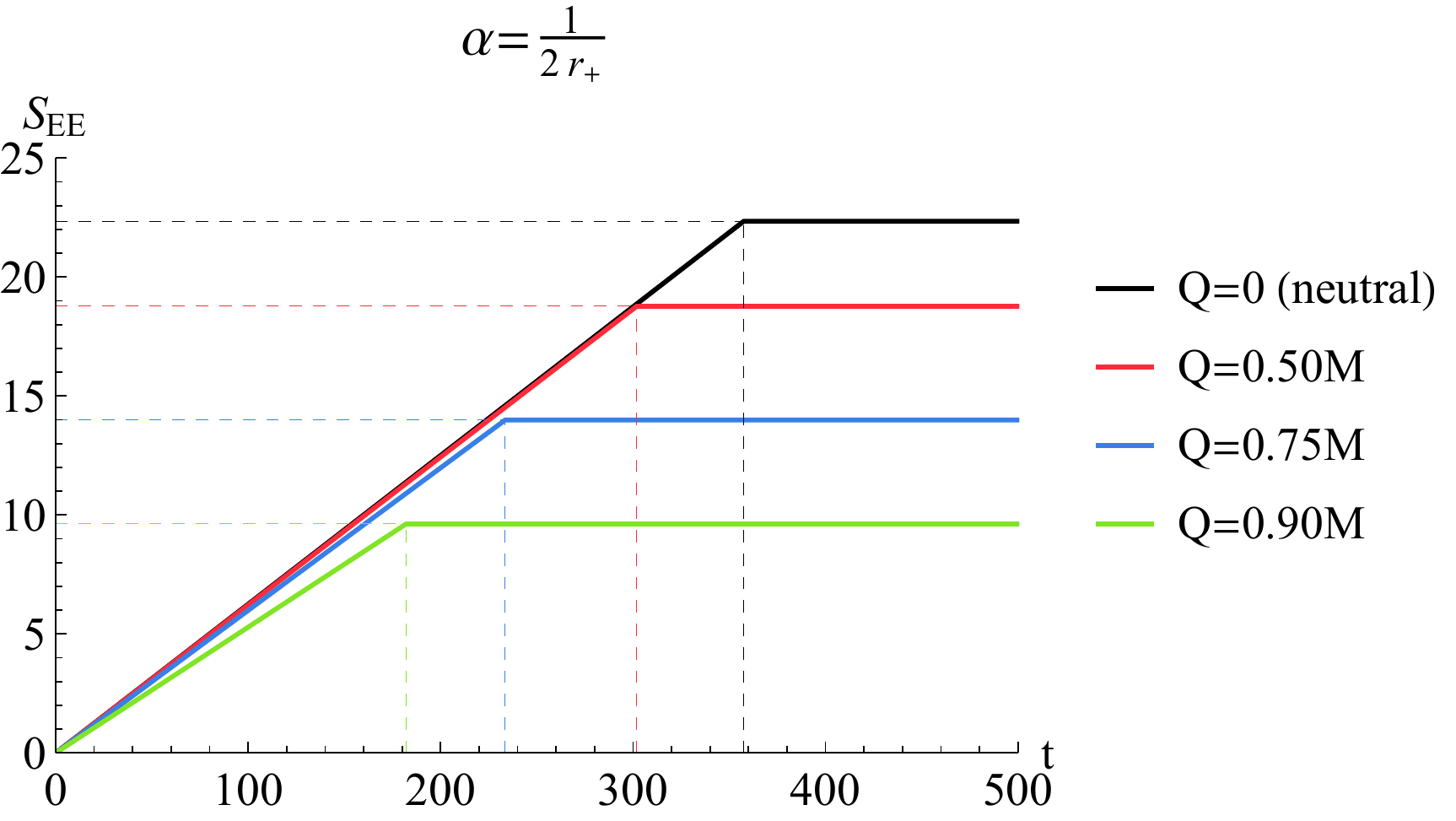}
}
\quad
\subfigure[\scriptsize{}]{\label{pc2}
\includegraphics[scale=0.4]{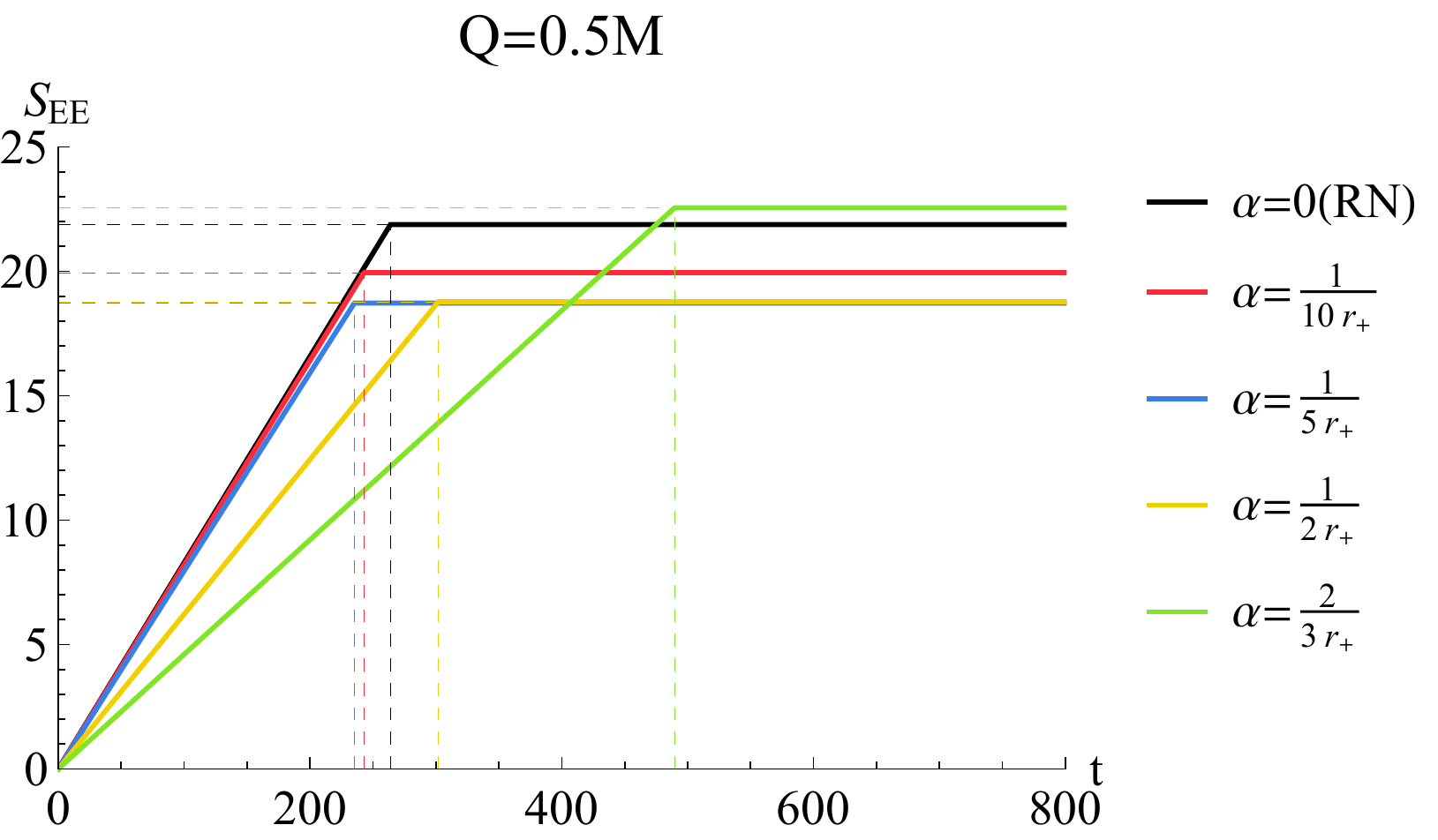}
}
\caption{The Page curve for eternal charged accelerating black holes, where setting $M=c=G_N=1$. On the left, we fix the acceleration $\alpha=\frac{1}{2r_+}$. On the right, we fix the charge $Q=0.5M$.}
\label{pagecurves}
\end{figure}

 Through these figures, one can easily find that in the $\alpha$-fixed case, the Page time increases as charges increase. But in the $Q$-fixed case, the Page time decreases first and then increases with the acceleration increasing. In particular, the Page time reaches its maximum for neutral black holes ($Q=0$). Note that these conclusions are applied only to non-extremal black holes. However, once the acceleration or charges makes the black hole approaches an extremal black hole, the Page time behaves divergent. Finally, the Page curves for non-extremal charged accelerating black holes are given in \mpref{pagecurves}.
\par
\begin{figure}[htb]
\centering
\subfigure[\scriptsize{}]{\label{scr1}
\includegraphics[scale=0.4]{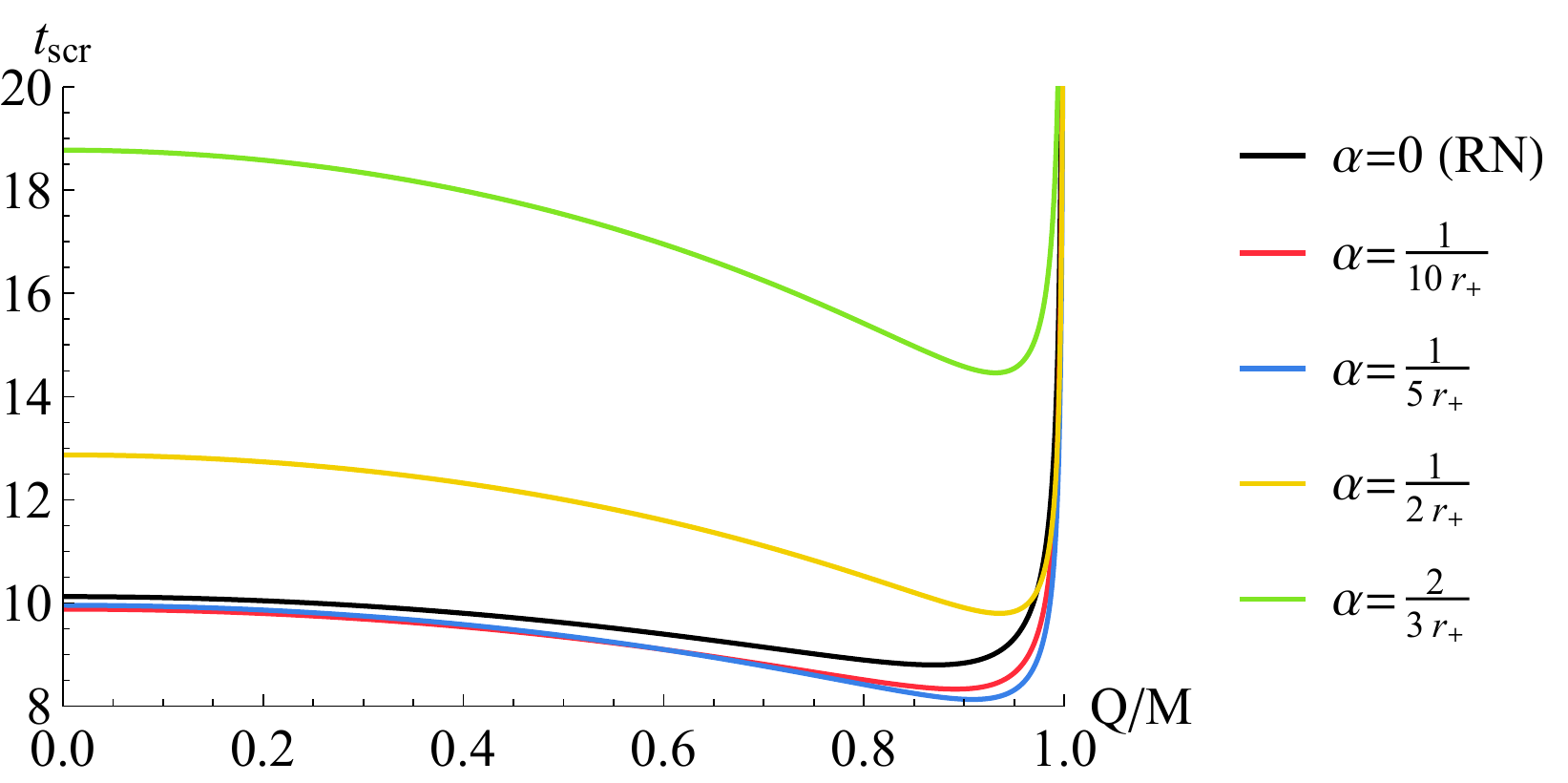}
}
\quad
\subfigure[\scriptsize{}]{\label{scr2}
\includegraphics[scale=0.4]{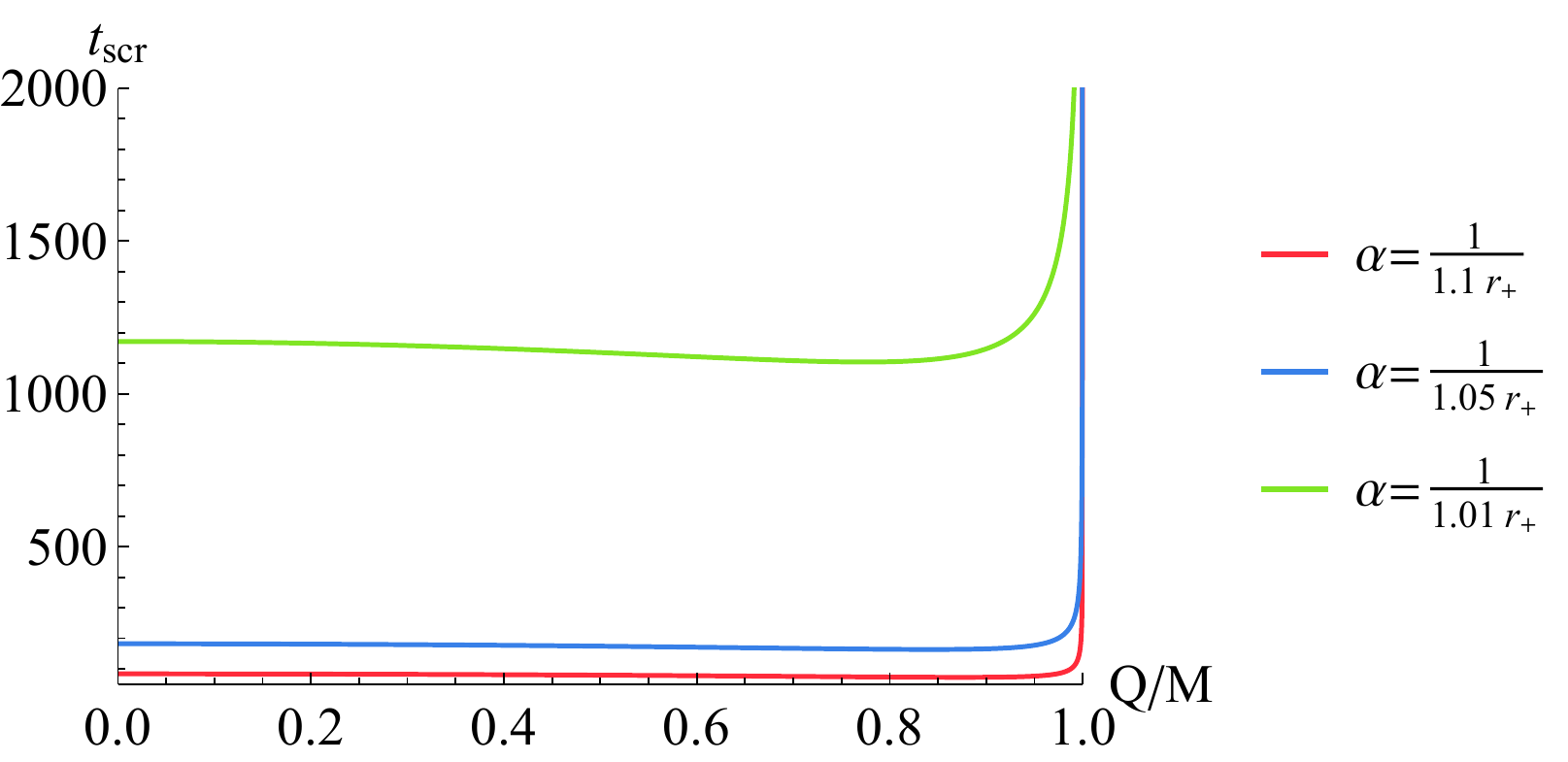}
}
\subfigure[\scriptsize{}]{\label{scr3}
\includegraphics[scale=0.4]{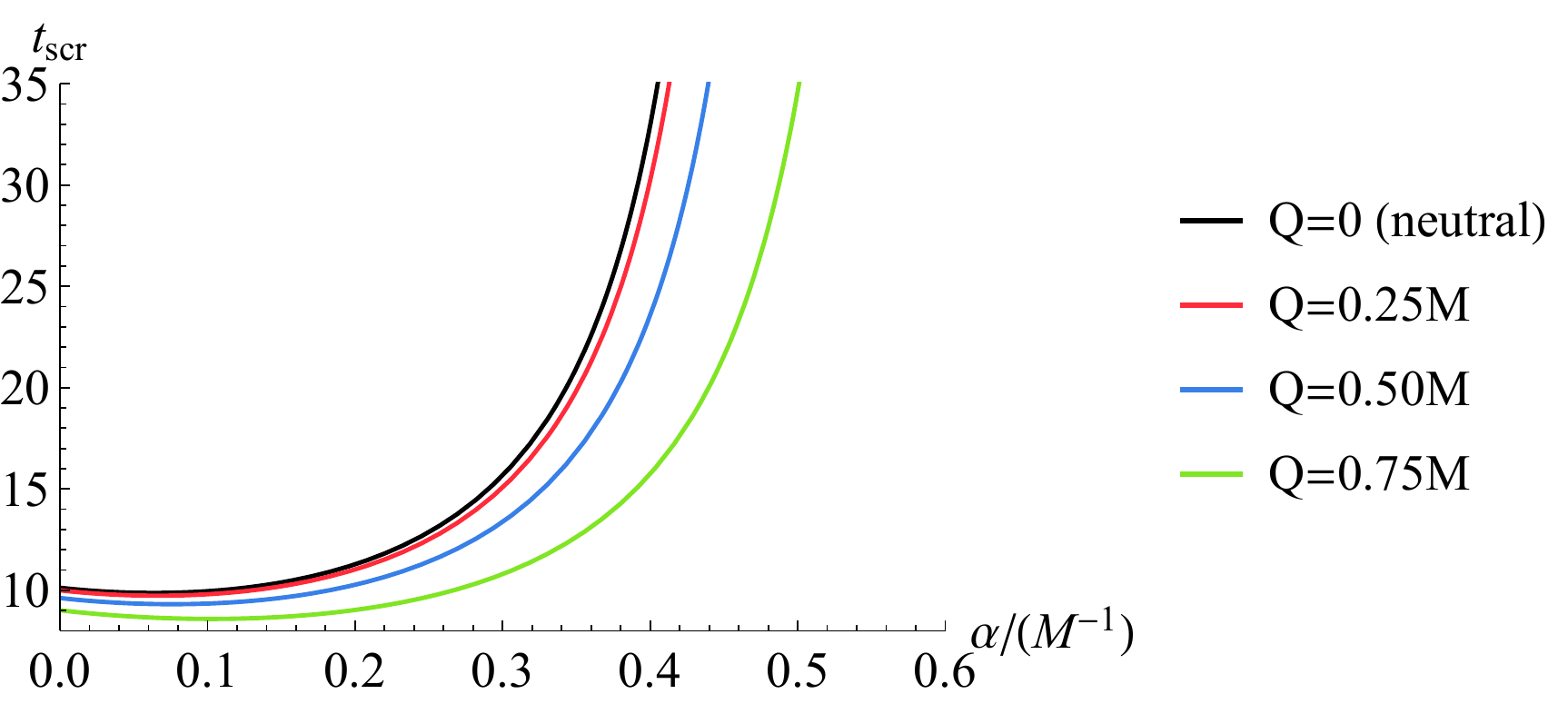}
}
\subfigure[\scriptsize{}]{\label{scr4}
\includegraphics[scale=0.4]{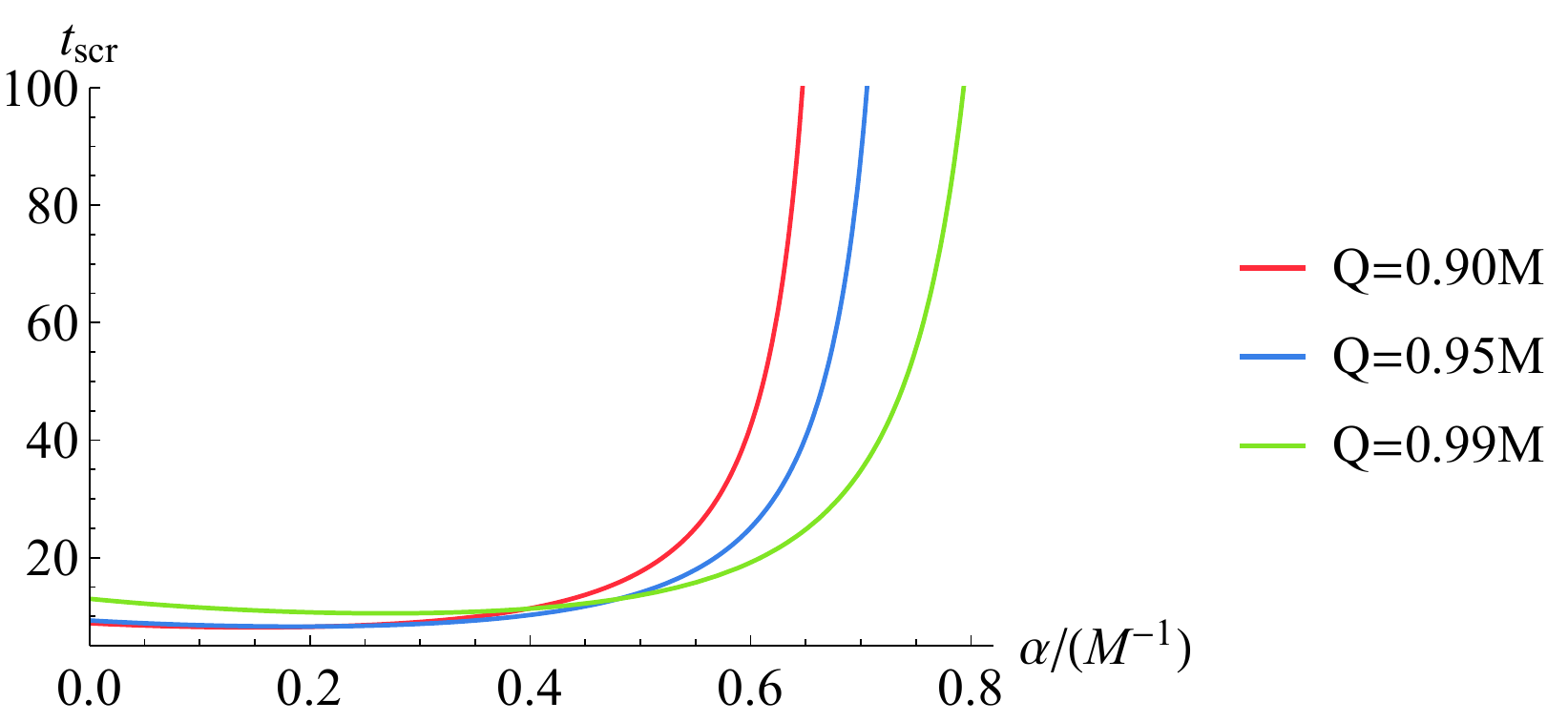}
}
\caption{The scrambling time by black holes as functions of charges $Q$ and accelerations $\alpha$. Here we set the mass $M=G_N=1$.}
\label{scramblingtime}
\end{figure}

 Now we discuss the scrambling time, which is based on the Hayden-Preskill Gedanken experiment: one drops a quantum diary into a black hole after the Page time. He would have to wait for what is so-called the scrambling time to recover this quantum information from Hawking radiation \cite{HP}. In the suggestion of the entanglement wedge reconstruction, the scrambling time is determined by how long it takes the information to touch the boundary of the entanglement island. Therefore, the asymptotic observer sends a light signal at the cutoff surface $(r=b)$, then the time when the signal reaches the island $(r=a)$ is
\begin{equation}
\text{v}(t_b,b)-\text{v}(t_a,a) =(t_b-t_a) + \big[ r_{\star} (b) -r_{\star}(a) \big],
\end{equation}
where $\text{v}(t,r)$ is denoted the null Kruskal coordinate \eqref{kruskal}. Thus, the scrambling time is the shortest time
\begin{equation}
\begin{split}
t_{\text{scr}} &\equiv t_b-t_a =\text{Min} \bigg \{ [r_{\star}(b) -r_{\star}(a)] -[\text{v}(t_b,b) -\text{v}(t_a,a)] \bigg \} \\
               &=r_{\star}(b) -r_{\star}(a).
\end{split}
\end{equation}
We obtain the expression of the scrambling time by invoking the location of the island \eqref{location}
\begin{equation}
\begin{split}
t_{\text{scr}} &=\frac{r_+^2}{2(r_+-r_-) (\alpha^2 r_+^2 -1) (\alpha^2 r_-^2 -1) }  \Bigg( (2\alpha^2 r_-^2 -2)  \log \bigg( \frac{a-r_+}{b-r_+} \bigg) + (2 \alpha^2 r_+^2-2) \log \bigg( \frac{b-r_-}{a-r_-} \bigg) \\
               &+\alpha (r_+-r_-) (\alpha^2 r_+ r_- +1) \bigg( \log \frac{1+b \alpha}{1-b \alpha}- \log  \frac{1+a \alpha}{1- a \alpha} \bigg) +(r_+^2 - r_-^2 )\alpha^2 \log \frac{1-\alpha^2 a^2}{1-\alpha^2 b^2}  \Bigg) \\
               &\simeq \frac{r_+^2}{2(r_+-r_-)(\alpha^2 r_+^2 -1)} \Bigg( 2 \log \frac{(a-r_+)(b-r_-)}{(b-r_+)(a-r_-)} +\alpha (r_+-r_-) \log \frac{(1-b\alpha)(1-a \alpha)}{(1-b\alpha)(1+a\alpha)} \Bigg)\\
               & \simeq  \frac{2r_+^2}{(r_+-r_-)(1-\alpha^2 r_+^2)} \log \frac{(b-r_+)(a-r_-)}{(a-r_+)(b-r_-)} \simeq \frac{1}{2 \pi T_H} \log \frac{r_+^2}{c G_N} \simeq \frac{1}{2 \pi T_H} \log S_{BH}.  \label{tscr}
\end{split}
\end{equation}
In the last line, we assume that $b$ has the order ${\cal O} (r_+)$ and the acceleration is not large $( \alpha \ll \frac{1}{r_+} )$. This result follows the conclusion of the initial Hayden-Preskill experiment \cite{scrambling time, acoustic}. Then, the scrambling time as a function of the acceleration $\alpha$  and the charge $Q$ is plotted in \mpref{scramblingtime}.
\par In the same way, the scrambling time decreases when the charge increases, while as the acceleration increases, it decreases and then increases. The neutral black hole also has the greatest scrambling time. However, in the extremal case, the scrambling time is also divergent. This is for the same reason that the Page time diverges in extremal cases. We will discuss some remarks in the next section.

\section{Conclusion and Discussion} \label{Summary}
\par In this paper, we study the information paradox in the charged ``$C$-metric'' that describes an accelerating black hole. Different for the RN black holes, the existence of acceleration parameters $\alpha$ results in accelerating black holes existing three horizons. Accordingly, the corresponding geometry is richer. Therefore, it is very meaningful to study the page curve under this background, and can further expand the scope of application of the island paradigm. We calculate the fine-grained entropy of the Hawking radiation explicitly by the island formula \eqref{final result}. Correspondingly, the Page curve is successfully reproduced by the island paradigm in \mpref{pagecurves}. In brief, the fine-grained entropy of radiation increases monotonically with time before the Page time. After this moment, it starts at a fixed value that is about two times of the Bekenstein-Hawking entropy. This is also consistent with the finiteness of the entanglement entropy for two-sided eternal black holes or a system at the thermofield double state.
\par A significant feature of the Page curve for the ``$C$-metric'' is the presence of the charge and the acceleration. In the non-extremal case, when they are both small, the increase in charges will cause the Page time to advance for fixed accelerations (\mpref{pagetime1}). On the contrary, the increase in acceleration will first advance and then delay the Page time (\mpref{pagetime2}). In addition, the dependence of the scrambling time on charges and accelerations has similar behavior to the Page time (\mpref{scramblingtime}). We also find that both the Page time and the scrambling time are maximized in neutral black holes. Note that, both the Page time and the scrambling time become divergent at the extremal case. A simple view is that the Hawking temperature of extremal black holes is vanishing, and the expressions of the Page time and the scrambling time are inversely proportional to the temperature. However, one needs to investigate extremal cases more prudently. Some reports suggest that the causal structure of extremal cases manifests different and the corresponding calculations need to be modified \cite{high31,high32}, while other work suggests the extremal black holes are completely dominated by the superradiance, which can cause a black hole to become into a neutral black hole, thus avoiding difficulties in extreme cases.  In this process, the Page time and the Bekenstein-Hawking entropy both decrease \cite{high21}. In brief, the Page curve of the extremal case remains an open question and more investigation is needed.
\par Another interesting open question involves the transfer of quantum information. The island formula just tells us an instruction on how to calculate the unitary Page curve, but the details of the escaped information falling into black holes are unknown. What happens when the black hole evaporates to a Planck scale, or does the wormhole appear at the end of the evaporation? Further studies into quantum gravity may be needed to uncover these intriguing questions.

\section*{Acknowledgement}
We would like to thank Yang Zhou, Rongxin Miao and Kilar Zhang for helpful discussions. The study was partially supported by NSFC, China (Grant No.12275166 and No.11875184).

\end{document}